\documentclass[twocolumn,showpacs,superscriptaddress]{revtex4}
\usepackage{here}
\usepackage{graphicx}
\graphicspath{{fig/}{}}

\usepackage{amsmath}
\usepackage{bm}% bold math

\newcounter{Fig}

\newcommand{\be}{\begin{equation}}
\newcommand{\ee}{\end{equation}}
\newcommand{\bea}{\begin{eqnarray}}
\newcommand{\eea}{\end{eqnarray}}

\begin{document}

%\twocolumn[ %% activate for two-column option

\title{All-optical switching and multistability in photonic structures with liquid crystal defects}

\author{Andrey E. Miroshnichenko$^{1,*}$, Etienne Brasselet$^2$, and Yuri S. Kivshar}

\address{Nonlinear Physics Center and Center for Ultra-high bandwidth
Devices for Optical Systems (CUDOS), \\Research School of Physical
Sciences and Engineering, Australian National University, Canberra
ACT 0200, Australia\\
$^2$Centre de Physique Optique Mol\'eculaire et Hertzienne, Universit\'e Bordeaux 1, CNRS,\\
351 Cours de la Lib\'eration, 33405 Talence Cedex, France\\
$^*$Corresponding author: aem124@rsphysse.anu.edu.au}

\begin{abstract}
We demonstrate that one-dimensional photonic crystals with pure
nematic liquid-crystal defects can operate as all-optical switching
devices based on optical orientational nonlinearities of liquid
crystals. We show that such a periodic structure is responsible for
a modulated threshold of the optical Fr\'eedericksz transition in
the spectral domain, and this leads to all-optical switching and
light-induced multistability. This effect has no quasi-statics
electric field analogue, and it results from nonlinear coupling
between light and a defect mode.
\end{abstract}

%\ocis{(190.4420) Nonlinear optics, transverse effects in; (190.5940) Self-action effects}

% ] %% activate for two-column option

\maketitle

The concept of photonic crystals~\cite{book} proposed two decades
ago~\cite{eli,sjohn} brought a new paradigm to achieve light
propagation control in dielectric media. In such periodic photonic
structures tuning may be achieved by using materials which are
sensitive to external fields, including temperature, electric field,
or light itself. In this context, nonlinear photonic crystals have
retained much attention due to possible enhancement of nonlinear
effects~\cite{soljacic}. Among various nonlinear optical materials
that can be implemented in actual photonic crystal devices, liquid
crystals (LCs) have been recognized as an attractive alternative
material~\cite{busch} due to their unique sensitivity to external
fields. Since then, many tunable photonic crystal devices based on
LC tunability  have been suggested and implemented either using
complete or partial LC infiltration into the periodic dielectric
structure. In the first case, the photonic bandgap is tuned due to
refractive index changes of the global structure~\cite{yoshino},
while in the second case a LC-infiltrated layer or hole generates
defect modes whose frequencies are controlled by local refractive
index changes of LC~\cite{ha01}. The case of complete infiltration
is the most studied one, and it was the first to be demonstrated; it
concerns thermal~\cite{yoshino} and electrical~\cite{kang}
tunability infiltrated one-, two- and three-dimensional photonic
structures with LCs. There exist much less studies concerning {\em
optical tuning}. One can mention the demonstration using photonic LC
fibers~\cite{PLCF04,all-optical_LC_07}, one-dimensional~\cite{CLC07}
or planar~\cite{Maune05,Kallassi07} photonic crystals using
absorbing or dye-doped LCs. In these works the resonant interaction
of light induces a change of the order parameter, phase transition or
surface-mediated bulk realignment. However the non-resonant case,
where well-known orientational optical nonlinearity of LC takes
place~\cite{zeldovich86}, has only been explored recently in
Ref.~\cite{miroshnichenko06}, where the optical Fr\'eedericksz
transition (OFT) was studied in a one-dimensional photonic crystal
with a nematic liquid crystal (NLC) defect. It was shown that a NLC
having a first-order OFT in the single slab case under linearly
polarized excitation could be used as an optical switch and operate
as an optical diode when the excitation wavelength matches a defect
mode frequency. However, such first-order OFT materials are not
common and the calculation was made with liquid crystal
PAA~\cite{miroshnichenko06}, which has the nematic phase in the
typical range of temperature $120 - 135^\circ$C and thus prevents
from applications in optical data processing.

% Figure 1
%
\begin{figure}[b]
\centering
\includegraphics[width=1\columnwidth]{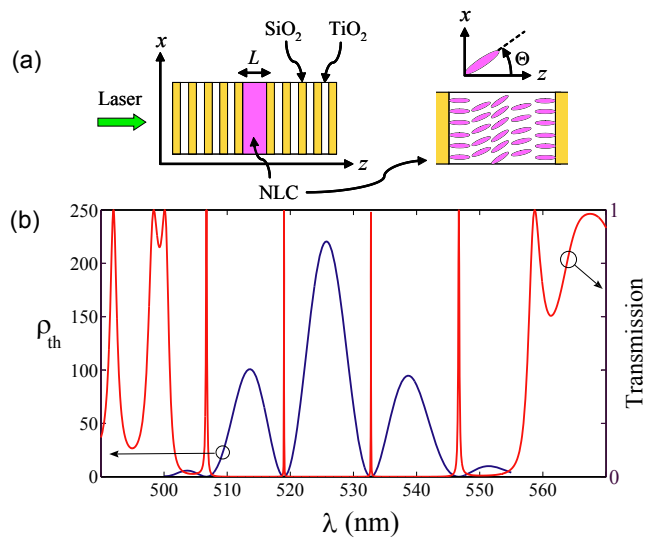}
\caption{(Color online) (a) Sketch of the multilayers periodic
structure with an embedded NLC defect. (b) Transmission spectra of
the unperturbed system (red) and normalized OFT threshold (blue).}
\label{fig:fig1}
\end{figure}

In this Letter, we study the effects of the orientational optical
nonlinearity of LCs in a one-dimensional periodic dielectric
structure for a linearly polarized input beam. First, we demonstrate
a nonlinear feedback due to coupling between light and a defect mode
via orientational nonlinearities. This feedback could be {\em
positive}, resulting in all-optical multistable switching, or {\em
negative}, depending on the wavelength detuning from a defect mode.
Second, we demonstrate that this coupling leads to different types
(first or second order) OFT in periodic structures for the same NLC
material. While all-optical bistable switches based on photonic
crystals are mainly restricted to inorganic
devices~\cite{all-optical_inorg_05}, our results could be envisaged
as a promising opportunity for LC infiltrated photonic crystals
technology.

The structure studied in this Letter is made of ten %alternating
bilayers of SiO$_2$ and TiO$_2$
%EB: can you confirm 'ten alternating bilayers', which is understood as 20 layers (ten SiO2 and ten TiO2). 'alternating' may be deleted because it is confusing since a bilayer already alternate materials... it is just superposition of bilayers...
on each side of a NLC layer having a
homeotropic alignment (molecules are perpendicular to substrates),
as shown in Fig.~1(a). The thicknesses of NLC, TiO$_2$ and SiO$_2$
layers are 5~$\mu$m, 134~nm  and 143~nm respectively. The resulting
structure has a bandgap centered around 530~nm that supports four
defect modes, as shown in Fig.~1(b) where the transmission spectra
of the unreoriented structure is presented. We choose the
commercially available E7 liquid crystal which is nematic at room
temperature and exhibits a second-order OFT for a single slab under
linearly polarized excitation.

The optical properties of the proposed structure are considered to
be within the plane wave approximation, so that the system depends
only on coordinate $z$ [Fig.~1(a)] and time $t$. The problem is
conveniently solved by using the Berreman $4\times4$ matrix
approach~\cite{berreman} where Maxwell's equations can be expressed
as $\partial{\mathbf\Psi}/{\partial z} = ik_0{\mathbf
D}{\mathbf\Psi}$, where $k_0=2\pi/\lambda$ is the wave vector in
vacuum, ${\mathbf D}$ is the Berreman matrix that depends on
dielectric permittivity tensor $\epsilon_{i\!j}$,
and ${\mathbf\Psi}=(E_x,H_y,E_y,-H_x)^T$~\cite{berreman}. 
The $E_z$ component of the
light field is obtained from the constitutive equation $\partial_i
\epsilon_{i\!j}E_{j}=0$. The whole structure is divided into many
layers with constant permittivity, which are described by constant
matrices ${\mathbf D}_n$. While the thickness of the dielectric
layers corresponds to thickness of actual materials, the liquid
crystal layer is discretized into as many sublayers as necessary to
ensure prescribed small variations from one sublayer to another. The
resulting matrix thus writes ${\mathbf D}=\prod {\mathbf D}_n$. We
obtain a light propagation problem for incident ($i$), transmitted
($t$) and reflected  ($r$) amplitudes that writes
${\mathbf\Psi}^{t}={\mathbf
D}({\mathbf\Psi}^{i}+{\mathbf\Psi}^{r})$. Inside the NLC layer, the
light field is coupled to the Euler-Lagrange equations that govern
the dynamics of the director ${\bf n}(z,t)$ [i.e. unit vector
associated with local averaged molecular orientation, see Fig.~1(a)]
and account for dissipative, elastic, and electromagnetic
contributions~\cite{zeldovich86}.

We further consider the propagation of light linearly polarized
along $x$-axis and assume the director to be constrained in the
$(x,z)$ plane, which can be represented by the reorientation angle
$\Theta(z,t)$ [Fig.~1(a)]. The problem is solved numerically via the
modal expansion procedure detailed in Ref.~\cite{etienne_JOSAB_CP}
by searching the reorientation profile as $\Theta(z,t) =
\sum_{n=1}^N \Theta_n(t)\sin(\pi z/L)$, where $N=10$ is enough to
ensure accurate results. We introduce the normalized intensity $\rho
= |E_x^i|^2/I_{\rm lin}^0$, where $I_{\rm lin}^0$ is the OFT
threshold for a single NLC slab and linearly polarized plane wave,
and time $\tau=t/\tau_{\rm NLC}$, where $\tau_{\rm NLC}$ is the
typical relaxation time~\cite{zeldovich86}.
Finally, we use 
\begin{eqnarray}\label{eq:Delta}
\Delta(\tau)=k_0\int_0^L(n_e(z,\tau)-n_o)dz\;,
\end{eqnarray}
with $n_{o/e}$ being the ordinary and effective extraordinary
refractive indexes, which is the total light-induced phase delay in
the presence of reorientation. All NLC material parameters are those
of Ref.~\cite{etienne_JOSAB_CP}, and the linear dispersion of
refractive indices are all taken into account.

% Figure 2
%
\begin{figure}[t]
\centering
\includegraphics[width=1.\columnwidth]{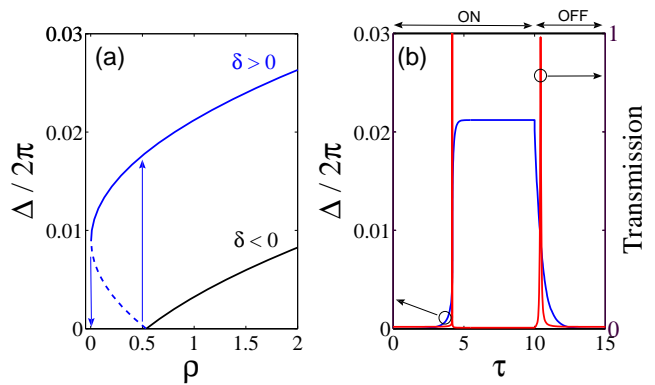}
\caption{(Color online) (a) Reorientation diagram near OFT for
positive (blue) and negative (black) detuning $\delta = \pm0.25$~nm
around $\lambda_{\rm d}=532.75$~nm. Solid (dashed) lines refers to
stable (unstable) states. (b) Corresponding dynamics, for the
$\delta>0$, of phase delay $\Delta$ and pump light transmission for
square-shaped temporal excitation with $\rho=1$.} \label{fig:fig2}
\end{figure}

First, we calculate the OFT threshold $\rho_{\rm th}$ above which
the NLC is reoriented for pump wavelength $\lambda_{\rm p}$ lying
inside the bandgap [see Fig.~1(b)]. As was shown in
Ref.~\cite{miroshnichenko06}, the reorientation threshold is
drastically reduced at a defect wavelength with respect to the single NLC
slab $\rho_{\rm th}(\lambda_{\rm d})\ll 1$, caused by a very
strong light confinement at the defect mode placed in a periodic
structure. A reduction factor up to $10^3$ is obtained for the
present structure. In contrast, a significant increase of the
threshold is observed when $\lambda_{\rm p}$ is strongly detuned
from defect mode frequencies [Fig.~1(b)]. Indeed, the smaller the
detuning parameter $\delta = \lambda_{\rm p}-\lambda_{ \rm d}$ is,
the better light confinement inside the defect NLC layer,
which leads to the lower reorientation threshold.

Although the OFT threshold is independent of the sign of $\delta$
for $\lambda_{ \rm p} \approx \lambda_{ \rm d}$ (local parabolic
approximation), the type of the light-induced reorientation strongly
depends on it.
This point is illustrated in Fig.~2(a) where the reorientation diagram
$\Delta$ as a function of $\rho$ calculated from (\ref{eq:Delta}) is shown for
$\delta=\pm0.25$~nm around $\lambda_{ \rm d}=532.75$~nm. A negative
detuning leads to a second-order OFT, as it is the case for the NLC
slab alone, whereas a positive detuning exhibits a first-order OFT
with an relative hysteresis width of almost 100\%. To understand
this behavior we note that defect mode frequencies underwent a
red-shift proportional to the increase of the averaged refractive
index inside the NLC defect layer due to molecular reorientation.
Consequently, recalling that the threshold intensity has a parabolic
profile around defect modes
$\rho_{\rm th}(\lambda_{\rm p}) - \rho_{\rm th}(\lambda_{\rm d}) \propto (\lambda_{\rm p}-\lambda_{ \rm d})^2$
, negative detuning $\delta<0$ leads to a
negative feedback and positive detuning $\delta>0$ is accompanied by
a positive feedback. All-optical bistable switching is thus achieved
by properly chosen pumping wavelength $\lambda_{ \rm p}>\lambda_{\rm
d}$.

The dynamics of the all-optical switching process is illustrated in
Fig.~2(b), where the unperturbed system is excited with $\delta>0$
and $\rho>\rho_{\rm th}(\lambda_{\rm p})$ for $\tau<10$ and $\rho=0$ for
$\tau>10$. Two transmission peaks are observed, the first one at
excitation stage and the second one during relaxation process. These
peaks represent red-shift and blue-shift of the defect modes
frequencies, respectively. Initial conditions at $\tau=0$ are taken
as $(\Theta_1=10^{-2},\Theta_{2,...,N}=0)$, which mimics a thermal
orientational fluctuation in the NLC slab.
As a result, the smaller the detuning parameter is, the smaller intensity is the required to
perform all-optical switching in periodic structures.
A compromise has nevertheless to be found to preserve light-induced phase delay discontinuity
since it vanishes for $\delta\to0^+$.

% Figure 3
%
\begin{figure}[t]
\centering
\includegraphics[width=1.\columnwidth]{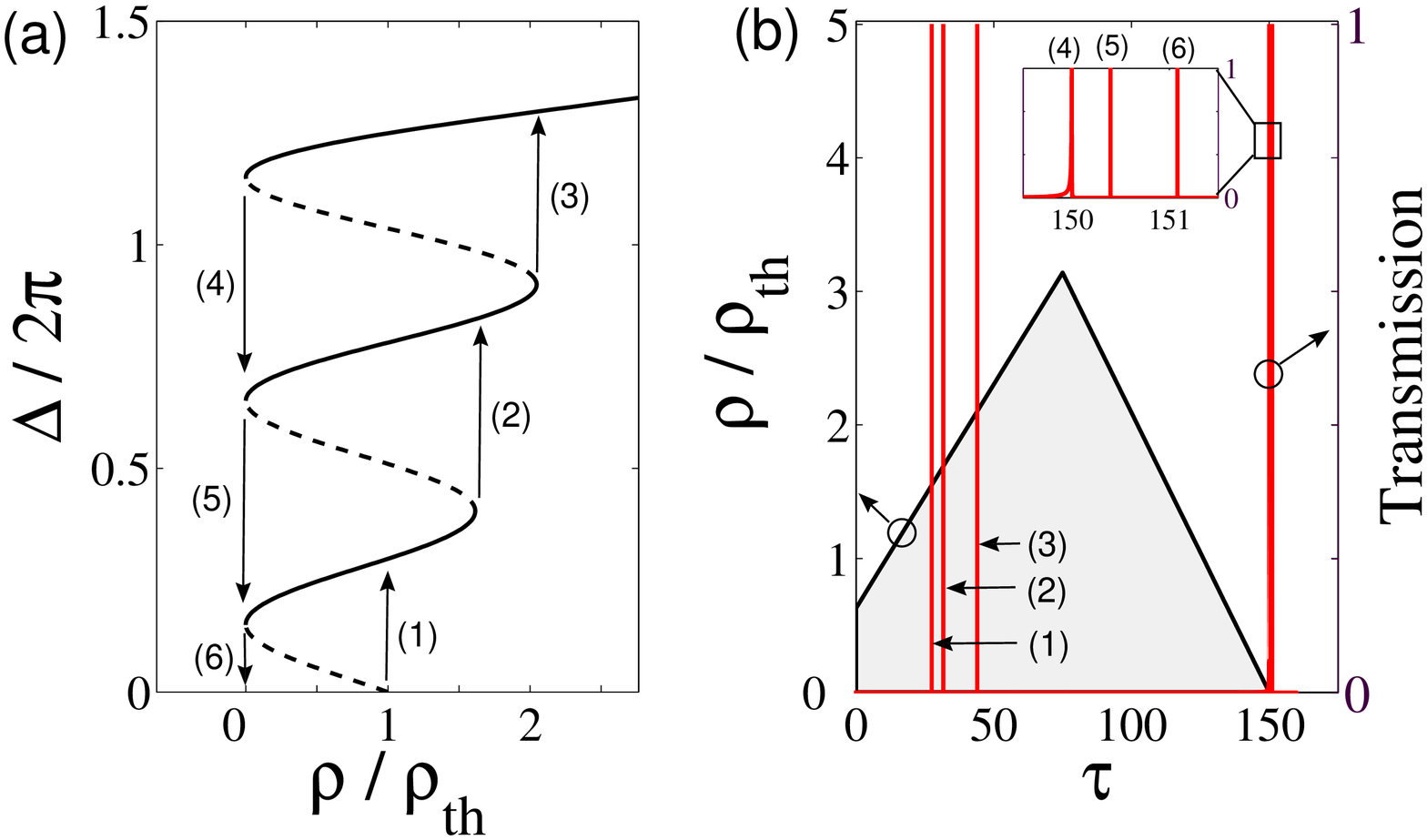}
\caption{(Color online) (a) Multistable reorientation diagram for
positive detuning at $\lambda_{\rm p} = 537$~nm. Solid (dashed)
lines refers to stable (unstable) states. (b) Corresponding dynamics
of pump light transmission for triangle-shaped temporal excitation.}
\label{fig:fig3}
\end{figure}

Multistable all-optical switching can also be achieved in such a
structure for larger intensities, caused by larger molecular
reorientation resulting in larger changes of the refractive index.
This leads to a possibility of many defect modes passing the pumping
wavelength $\lambda_{\rm p}$, and, consequently, to a series of
reduction of the pumping intensity $\rho$ giving reorientation boosts.
An example is shown in
Fig.~3(a) for $\lambda_{\rm p} = 537$~nm. The multistability is
evidenced by several coexisting stable states in the reorientation
diagram. The corresponding switching dynamics is shown in Fig.~3(b)
when a linear ramp of excitation light intensity is used. Each of
six peaks of the transmission dynamics is a reminiscent of a plateau
in the reorientation diagram, as indicated by labels numbered (1-6).

In conclusions, we have predicted multistable all-optical switching
resulting from an unique nonlinear coupling between light and defect
modes in a periodic dielectric structure that contain a NLC defect layer. 
Moreover, depending on the pumping
wavelength $\lambda_{\rm p}$ the same NLC material may demonstrate
first- or second-order OFT in periodic structures. In contrast to
other optically tunable photonic structures based on liquid
crystals, the proposed system is based on orientational optical
nonlinearity of liquid crystals. It is important to note that there
is no quasi-statics electric field analog to the optical case
studied here because the electro-optical tunability explored
earlier~\cite{ozaki07} is not accompanied by a coupling between
excitation field and defect modes. Additional functionalities are
expected when non-planar molecular light-induced reordering takes
place~\cite{etienne_JOSAB_CP}.

This work was supported by Discovery and Center of Excellence
projects of the Australian Research Council.

\end{document}